\documentclass[
    ,final            
  ]
  {aipproc}
\usepackage{wrapfig}
\layoutstyle{6x9}

\begin{document}

\title{The Leptonic Higgs and Dark Matter}


\author{Piyush Kumar}{
  address={Berkeley Center for Theoretical Physics, University of California,
Berkeley, CA 94720\\Theoretical Physics Group, Lawrence Berkeley National Laboratory, Berkeley, CA 94720}
}

\begin{abstract}
This is a short review of the framework proposed in \cite{Goh:2009wg} which gives rise to indirect Dark Matter (DM) 
signals explaining the recent cosmic-ray anomalies and links cosmic-ray signals of DM to LHC signals 
of a leptonic Higgs sector.The states of the leptonic Higgs doublet are lighter than about 200 GeV, yielding large
$\bar{\tau} \tau$ and  $\bar{\tau} \tau \bar{\tau} \tau$ event rates at the LHC. For the case of
annihilations, cosmic photon and neutrino signals are constrained.
\end{abstract}

\maketitle


\section{Introduction}

Recent observations of high-energy electron and positron cosmic ray spectra
have generated tremendous interest in the astrophysics and particle physics community.
The PAMELA experiment reported an excess of positrons in
the few GeV to 100 GeV range. In
addition, results from the FERMI-LAT and HESS experiments suggest an excess of electrons and
positrons in the 100 GeV to 1 TeV range.  

The cosmic-ray data from these experiments show some puzzling features. For example, PAMELA suggests an excess only in positrons, but not in antiprotons. In addition, although FERMI and HESS experiments seem to suggest an excess in electrons and positrons compared to the naive astrophysical background, the excess is not as sharp as that reported by the earlier results of ATIC. There is still a considerable amount of uncertainty in the astrophysical processes which provide the background for the above signals. Therefore, the explanation of these observations imply one or more of the following : a) Our present understanding of the astrophysical background is incomplete, b) There is a nearby astrophysical source of primary electrons and positrons, such as pulsars, and c) The source of primary electrons and positrons is Dark Matter (DM) whose interactions are governed by an underlying particle physics model.

\section{The Framework}

In \cite{Goh:2009wg}, we focussed on c) as an explanation of the cosmic-ray data. Any particle physics model of DM trying to explain the excesses in the above experiments must address two important issues. First, for annihilating DM these signals require that the
annihilation cross-section for DM particles is typically two to
three orders of magnitude larger than that expected from the thermal
freezeout of WIMP DM. The discrepancy between the two cross-sections can, however, be resolved in scenarios which contain  
``moduli" whose decay could give rise to non-thermal production of Dark Matter and also in scenarios which contain  
light particles in the dark sector giving rise to a Sommerfeld enhancement in the \emph{present} cross-section. For decaying DM, the relic abundance could be given by conventional thermal freezeout, but the life-time of the DM particles must be extremely large ($\sim 10^{25}$-$10^{26}$s) and new physics is needed to explain it. As we will see later, this can also be achieved in a natural manner. Second, the signals apparently require
annihilations or decays dominantly into leptons rather than hadrons, since there is no reported excess in anti-proton cosmic rays (assuming that our understanding of the astrophysical background is roughly correct). There are two natural ways to achieve this - i) by {\it kinematics}, since leptons are lighter than quarks, and ii) by a {\it symmetry}, which forces the DM to annihilate/decay dominantly to leptons. 

In addition to the known physics of Standard Model (SM) particles and their interactions, we expect new physics to include both a Higgs sector, responsible for electroweak symmetry breaking, and a DM sector. The idea of a WIMP DM sector is particularly interesting as it ties in with general ideas about electroweak symmetry breaking. 
\begin{wrapfigure}{r}{0.4\textwidth}
 \vspace{-10pt}
  \begin{center}
     \resizebox{2.5in}{!}{\includegraphics*[0,730][338,789]{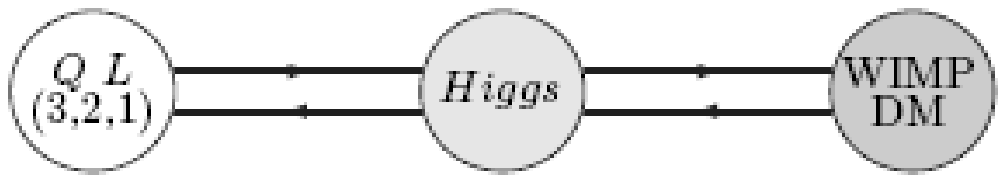}}
  \end{center}
  \vspace{-10pt}
\end{wrapfigure}

The interactions between the higgs sector and the Standard Model sector are well known; on the other hand the interactions of the WIMP DM sector with the other two sectors are more speculative. However, the WIMP idea is that the mass scales of
the dark matter and Higgs sectors are \emph{both} related to the weak scale, and this strongly
suggests some connection between these two sectors. With the above philosophy, in \cite{Goh:2009wg} we studied a general framework exploring the possibility that any direct couplings between the WIMP sector and the visible sector are subdominant; thus the Higgs sector is seen to be the messenger that makes the WIMP visible to us, as shown in the schematic  above. The nature of the Higgs sector is determined by the constraints from the cosmic-ray data. The Higgs sector in the SM or the MSSM does not give rise to a good fit to the data as in these cases the higgs decays predominantly to pairs of $W$ bosons, bottom quarks or top quarks (if heavy enough), giving rise typically to too many anti-protons than actually observed by PAMELA. Thus, from the data we are quite generally led to a framework in which the higgs couples dominantly to leptons, hence the name ``leptonic higgs". Such a higgs sector can be naturally obtained, for example by imposing an appropriate approximate $Z_2$ symmetry on the general two-higgs doublet model\cite{Goh:2009wg}. Many classes of models for the Dark sector, utilizing both annihilations and decays, can be constructed within this general framework, demonstrating its generality. 

Since the higgs couples via yukawa couplings, the leptonic higgs will dominantly decay to pairs of tau leptons which will subsequently cascade to electrons and positrons. A straightforward consequence of this is that fitting the PAMELA data requires a larger DM mass than in models where the WIMP cascades directly to electrons or muons, and that the FERMI and HESS signals are quite smooth. Also, the signal for the positron fraction in PAMELA is expected to continue to energies larger than 100 GeV, and not to show a very sharp peak. Another very interesting feature of the framework results from the production of tau leptons. Since decays of taus give rise to an ${\cal O}(1)$ fraction of photons (mostly from $\pi^0$s coming from tau decays) and neutrinos (from three-body tau decays), there is an exciting possibility of detecting energetic gamma rays and neutrinos in the energy range 100-1000 GeV. However, these signals depend on whether the taus originate from DM annihilations or DM decays.

\section{SIGNALS}

\begin{figure}[htp]
      \resizebox{8cm}{!}{\includegraphics[0,0][494,317]{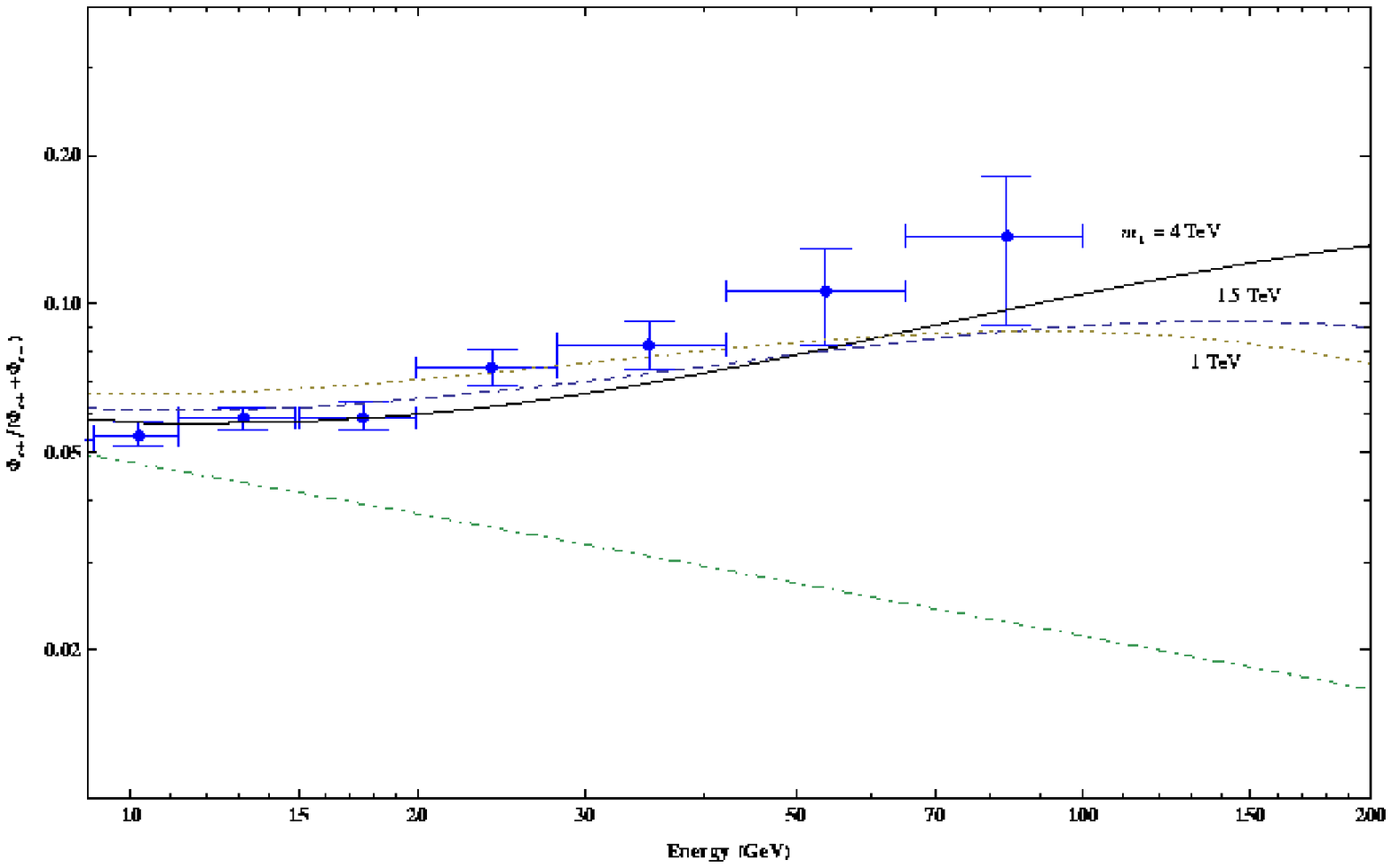}} \hfill 
      \resizebox{8cm}{!}{\includegraphics[13,14][829,541]{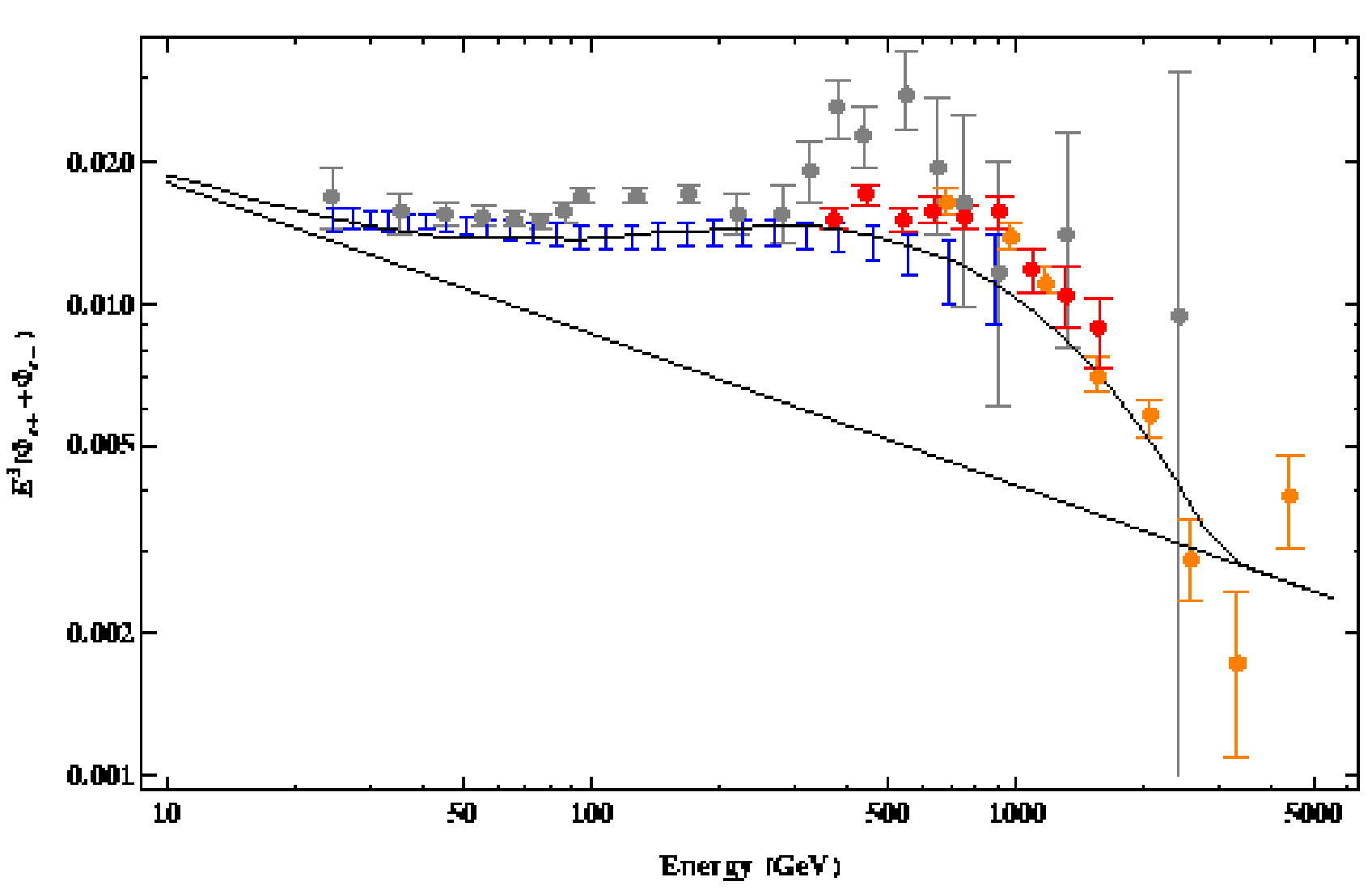}}
\caption{{\bf Left}: The positron fraction for 1, 1.5 and 4 TeV DM masses for the ``$4\tau$" annihilation mode with $B_e^{tot}$ given by 1200, 1950 and 10000 respectively. {\bf Right}: Results for $E^3\,(\Phi_{total}^{e+}+\Phi_{total}^{e-})$ for a 4 TeV DM mass with $B_e^{tot}=10000$ for the same mode. The MED propagation model is used. The average local halo density is taken to be $0.3 {\rm GeV/cm^3}$.}
\label{pamfermi}
\end{figure}

The main focus of \cite{Goh:2009wg} was to emphasize the general nature of the Higgs sector through which the WIMP DM sector couples to the visible sector, and not to specialize to a particular DM sector. Therefore, both annhilation and decay modes of the DM particles were studied and simple models giving rise to both modes were provided. Also, depending on whether DM is a scalar or fermion, the leptonic cosmic ray signals could arise from a variety of channels: DM annihilation to $\tau^4$, $\tau^2\nu^2$ or DM decay to $\tau^4,\tau^2,\tau^2\nu,\tau\nu l$ via intermediate higgs states \cite{Goh:2009wg}. Since the annihilation and decay modes can be related in a simple manner, results are shown  in Figure \ref{pamfermi} for a particular annihilation mode in which the DM annihilates to 4$\tau$'s via the higgs scalar $H$ and the psuedoscalar $A$. Good fits to both PAMELA and FERMI, HESS can be obtained for $m_{DM}$ around 4-5 TeV with a \emph{total} electronic boost factor $B_e^{tot}$ of about 10,000. The boost factor is an enhancement factor which could arise from many sources, such as a larger cross-section compared to the ``standard" thermal one, a clumpiness in the local DM halo, etc.  The precise spectrum depends on the astrophysical parameters of the propagation model of electrons and positrons with a mild dependence on the masses of the intermediate higgs states; hence the above masses and electronic boost-factors should only be taken as an estimate. 
      
In order to test approach, it is useful to look for \emph{correlated} signals. As mentioned earlier, production of taus will inevitably give rise to a large number of energetic photons and neutrinos arising from the decay of the taus. Since the signal from annihilations depends on the square of the DM density  $\rho_{DM}^2$ in contrast to $\rho_{DM}$ for decays, the constraints from the non-observation of photons and neutrinos from the Galactic Center (GC) and nearby dwarf galaxies are much more stringent for annihilations than for decays. For example, for the annihilation channel the constraints from photons can typically only be satisfied for shallow DM profiles and for a smaller photon boost-factor $B_{\gamma}^{tot}$ than the electronic one $B_e^{tot}$. At the same time, however, the annihilation modes also lead to better detection possibilites for future experiments. For the decay modes, it turns out that neutrinos provide better detection prospects than photons \cite{Goh:2009wg}.

One of the most interesting aspects of this framework is the existence of not only correlated cosmic-ray signals, but also correlated Higgs signals at the LHC! Since the DM masses typically turn out to be a TeV or higher, it is hard to produce the DM particles directly at the LHC but the leptonic and hadronic Higgs states could be produced at the LHC since they are ${\cal O}(100)$ GeV, which subsequently decay to tau leptons. For concrete predictions, we studied a two-higgs doublet model in which a softly broken discrete symmetry (parity) forces one of the higgs to couple dominantly to leptons and the other to quarks. This gives rise to extremely interesting 4$\tau$ signals at the LHC from Drell-Yan production of the leptonic higgs and its pseudoscalar partner,  followed by their decays to tau pairs. The same process also gives rise to production of charged higgs pairs $H^{\pm}$ which dominantly decay to $\tau^{\pm}\nu$. This channel therefore provides a new search strategy for the Higgs which has not been well studied so far, and could provide a discovery channel for modest luminosities around 30 $fb^{-1}$.  In addition, we find that the $2\tau$ signal from single higgs (both leptonic and hadronic) production by vector boson fusion followed by decay to tau pairs can be naturally enhanced compared to that for the Standard Model Higgs, and hence could also provide a discovery channel at modest luminosities. So, Higgs physics seems to be extremely promising.

Finally, we comment on the general compatibility of DM models trying to explain cosmic ray signals, with the observed upper bound on the DM relic abundance. For annihilating DM, the cross-section required to explain the cosmic-ray signals is typically larger by $\sim 100-1000$ than that required to obtain the correct relic abundance from a ``standard" thermal freeze-out computation. Since the (leptonic) higgs messengers of DM are naturally of ${\cal O}(100)$ GeV, there is no large sommerfeld enhancement as in models proposed by \cite{ArkaniHamed:2008qn}. However, the correct relic-abundance can be obtained even for a much larger annihilation cross-section if there exist late-decaying light scalar fields (moduli)  because then the relic abundance is determined by the reheat temperature of the moduli rather than the freezeout temperature of DM. This leads to non-thermal production of DM, which with some reasonable assumptions, surprisingly has the correct abundance, due to the existence of a non-thermal WIMP ``miracle" as emphasized in \cite{Acharya:2009zt}. In fact, such light moduli fields automatically occur within ``realistic" string theory compactifications, hence non-thermal production of DM is quite natural \cite{Acharya:2008bk}. For decaying DM, one has to explain the extremely long lifetime ($\sim 10^{26}$s) required for the cosmic ray data. It turns out that if the lepton parity (for a leptonic higgs) and dark parity (the parity which keeps the DM stable) are spontaneously broken by $\sim v_{EW}/M_{GUT}$ , then a lifetime of the correct magnitude can be naturally obtained. Note that signals for LHC Higgs physics are the same for both annihilation and decay modes since the parity breaking effects are extremely small.

\end{document}